\documentclass[a4paper,11pt]{article}

\usepackage{jheppub} 

\usepackage[T1]{fontenc} 

\newcommand{\be}{\begin{equation}}
\newcommand{\ee}{\end{equation}}
\newcommand{\bea}{\begin{eqnarray}}
\newcommand{\eea}{\end{eqnarray}}
\newcommand{\bse}{\begin{subequations}}
\newcommand{\ese}{\end{subequations}}
\newcommand{\bi}{\begin{itemize}}
\newcommand{\ei}{\end{itemize}}

\newcommand{\nn}{\nonumber}

\def\s2s1{S$^2\times$S$^1$ }
\def\Label#1{\label{#1}%
  \smash{\hbox to0pt{\raise1ex\hbox{\tiny[#1]}\hss}}}
\def\noLabels{\let\Label=\label}
\def\nobbibitem{\let\bbibitem=\bibitem}

\newcommand{\rd}{\partial}

\title{\boldmath More on the Entropy Product and Dual CFTs}
\author[a,1]{Hanif Golchin,\note{h.golchin@uk.ac.ir}}
\affiliation[a]{Faculty of Physics, Shahid Bahonar University of Kerman, \\
PO Box 76175, Kerman, Iran.}

\abstract{In this paper, we check the universality of the entropy product for some black hole and black ring solutions. Applying the asymptotic symmetry group (ASG) analysis, we also find the central charges of the dual CFTs for these solutions. It has been observed \cite{Chen:2013rb} that if the entropy product for a solution is universal, it is possible to read the central charges of the dual CFTs from the entropy product as $c_i \sim \frac{\partial}{\partial N_i}(S_+S_-)$, where $N_i$ is a conserved charge of the solution like angular momentum or electric charge. In this work we consider some other families of solutions and we check that the same behavior is valid for them. Moreover, in the case of solutions containing the conical singularity, we find the central charges using the ASG analysis. We show that the central charges receive a contribution from the conical characteristic $\kappa$ of the solution, however it is still possible
to read the central charges from the entropy product.}
 

\begin{document}
\maketitle
\flushbottom

\section{Introduction}
For more than 45 years, the quantum aspect of black holes is one of the most interesting and challenging areas in the theoretical physics. The brilliant achievements such as the Hawking radiation \cite{Hawking:1974sw}, black hole microstates counting \cite{Strominger:1996sh} and AdS/CFT correspondence \cite{Maldacena:1997re}, inspired people to  explore the relations between quantum field theories and black hole solutions. In this way, one of the progresses is the Kerr/CFT correspondence \cite{Guica:2008mu} which proposes that the quantum states in the near-horizon of an extremal Kerr black hole are identified with a two-dimensional chiral conformal field theory (CFT). This CFT arises from the generators of the asymptotic symmetry group (ASG) of a certain boundary conditions which is imposed on the excitations of the near horizon extremal Kerr geometry. Since the Kerr/CFT correspondence is extended to many rotating black hole and black ring solutions \cite{Loran:2008mm}-\cite{Chen:2013rb}, in the following we call them rotating solutions/CFT (Rot.Sol./CFT) correspondence.

 Thermodynamics of black hole horizons and the entropy product for multi-horizon black holes are also received considerable attentions in the recent years \cite{Chen:2012mh}-\cite{Golchin:2019hlg}. It has been observed that the product of horizon entropies for many solutions, is mass independent as $S_+S_-=(2\pi)^2N$\,, where $N$ depends only on the quantized charges like angular momentum and electric charge. This property is called ``universality'' of the entropy product \cite{Ansorg:2009yi,Ansorg:2010ru}. The mentioned universality is also observed in the higher dimensional black hole and black ring solutions \cite{Cvetic:2010mn,Castro:2012av}. 
  It has been also shown \cite{Golchin:2019hlg} that in the case of black hole and black ring solutions with conical singularity, the entropy product is mass independent, however the conical characteristic $\kappa$ of the solutions, appears in the universality relation as $\kappa S_+S_-=(2\pi)^2N$\,.

It is worth mentioning that  the entropy product $S_+S_-$ is universal (mass independent) provided that $T_+S_+=T_-S_-$\,, where $T_{+}$ and $T_-$ are Hawking temperatures on the outer and inner horizons, respectively \cite{Chen:2012mh}. 
The thermodynamics of black hole horizons and its relation with the Rot.Sol./CFT correspondence is also studied. In fact, it is proposed in \cite{Chen:2012mh,Chen:2012yd,Chen:2013rb} that when the entropy product is universal for a solution, one may reads the central charges of the dual CFTs from the universality relation as \cite{Chen:2013rb} $c_i =\frac{6}{4\pi} \frac{\partial}{\partial N_i}(S_+S_-)$, where $N_i$ is a conserved charge of the solution as spin and electric charge (all conserved charges except mass). Note that in the Rot.Sol./CFT analysis one finds the central charges of the dual CFTs only at the extremal limit however, in the thermodynamical approach presented in \cite{Chen:2013rb}, it is possible to find central charges for the generic (non-extremal) solutions.

In this work using the ASG analysis, we will find the central charges of the dual CFTs for five black hole and black ring solutions. We also check the universality of the entropy product for these solutions. We find that the thermodynamical method proposed in \cite{Chen:2013rb} is  true for the regular (without conical singularity) solutions. In the case of solutions containing the conic singularity, we find out that the central charges receive a contribution from the conical characteristic $\kappa$ of the solutions.  However, one can still obtain the correct central charges from the thermodynamics method of \cite{Chen:2013rb}.

The organization of this paper is as follows. In section \ref{s2}, we briefly review some examples in which the central charges of the dual CFTs are found by using the thermodynamical method \cite{Chen:2013rb}. In section \ref{s3}, for two regular (without conic singularity) solutions, we calculate the central charges of the dual CFTs using the ASG analysis. We also demonstrate the universality of the entropy product and find the central charges by the method proposed in \cite{Chen:2013rb}. In section \ref{s4}, we investigate three black hole and black ring solutions with conical singularity, using the approach of section \ref{s3}. We show that the method of \cite{Chen:2013rb}  is also valid for the solutions with conic singularity. Section \ref{s5} is devoted to the concluding remarks.

\section{Entropy products and central charges} \label{s2}
In this section, we review the relation between the horizons area product and central charges of the dual CFTs for some black hole and ring solutions. Throughout the paper we set the Newton's gravitational constant to $G=1$.

\vspace{2mm}
{\bf The Kerr-Newman black hole}\\ The four dimensional Kerr-Newman black hole \cite{Newman:1965my} is characterized by its three conserved charges: mass ($M$), spin ($J$) and electric charge ($Q$). It has been shown \cite{Ansorg:2009yi} that for this solution the entropy product of inner and outer horizons is universal as
\be \label{unikn}
S_+S_-=4\pi^2 \Big(J^2+\frac{Q^4}{4}\Big)\,.
\ee
Applying the Rot.Sol./CFT analysis for this solution, the central charge of the dual CFT can be obtained (in the J-picture) as \cite{Chow:2008dp,Chen:2011wm}
\be \label{jkn}
c^J=12\,J\,,
\ee
it is also possible to find the Q-picture central charge as \cite{Chen:2011wm}
\be \label{qkn}
c^Q=6\,Q^3\,.
\ee
On the other hand, it is proposed in \cite{Chen:2012mh,Chen:2012yd,Chen:2013rb} that if the area product of the horizons is universal for a solution, one can find the central charges of the dual CFTs as 
\be \label{unigen}
c_i=\frac{6}{4\pi^2}\,\frac{\partial (S_+S_-)}{\partial N_i}\,,
\ee
 where $N_i$\,s are the conserved charges of the solution like angular momenta and electric charges. In fact, using (\ref{unigen}) and considering the universality relation (\ref{unikn}), one may read the central charges as
\be
c^J=6\,\frac{\partial}{\partial J}\,\Big(J^2+\frac{Q^4}{4}\Big)=12J\,, \qquad c^Q=6\,\frac{\partial}{\partial Q}\,\Big(J^2+\frac{Q^4}{4}\Big)=6\,Q^3\,,
\ee 
which are in agreement with (\ref{jkn}) and (\ref{qkn}). In  other words, universality of the entropy product in a solution is a hint for the existence of the dual CFTs\footnote{Note that the formalism of \cite{Chen:2012mh,Chen:2012yd,Chen:2013rb} works for the generic (non-extremal) solutions, while the Rot.Sol./CFT analysis is valid in the case of extremal solutions.}.

\vspace{2mm}
{\bf Neutral double rotating black ring}\\
The horizon topology of black rings in five dimensions is $S^1\times S^2$. It is discussed in \cite{Castro:2012av} that the product of  horizons entropy may be universal, regardless of the topology of horizons. In the case of the five dimensional neutral double rotating  black ring \cite{Pomeransky:2006bd}, the solution is characterized by its mass ($M$) and two angular momenta $J_{\phi}, J_{\psi}$. It is observed \cite{Castro:2012av} that the entropy product in this case satisfies the universality relation
 \be \label{uni2s}
 S_+S_-=4\pi^2J_{\phi}^2\,.
 \ee
It is worth mentioning that in this solution, the $S^1$ circle of the black ring is parametrized by the $\psi$ coordinate and $J_{\phi}$ is the angular momentum perpendicular to it. In the thermodynamical method \cite{Chen:2013rb}, using (\ref{unigen}) one can read the central charges of dual CFTs as
\be
c_{\phi}=12J_{\phi}\,,\qquad c_{\psi}=0\,,
\ee
and the same result is obtained by doing the Rot.Sol./CFT analysis \cite{Chen:2012yd,Ghodsi:2013soa}. 

 The absence of CFT dual to the rotation along $\psi$ is explained in the thermodynamical method by the absence of $J_{\psi}$ in the entropy product (\ref{uni2s}). While in the Rot.Sol./CFT analysis, it can be explained as follows. The quantum states in the near horizon of the {\it extremal} solution, are identified with a certain two dimensional CFT (as mentioned in the introduction). In other words, the {\it extremality} plays a key role in this analysis. In the case of five dimensional single rotating neutral black ring (with the angular momentum $J_{\psi}$ along the circle $S^1$), the solution does not admit the smooth extremal limit \cite{Sadeghian:2015hja}. However the extremality may happen by adding the second rotation $J_{\phi}$. More explicitly, the extremality comes from the second rotation $J_\phi$ and therefore, there is just one CFT corresponding to the rotation along the $\phi$ direction.

\vspace{2mm}
{\bf Single rotating dipole black ring}\\
 This solution \cite{dipole}, contains dipole charge $q$ in addition to an angular momentum $J_{\psi}$ along the $S^1$ circle of the ring. The universality of the entropy product is also observed for the single spin dipole black ring as \cite{Castro:2012av,Chen:2012yd}
\be
S_+S_-=4\pi^2J_{\psi}q^3\,.
\ee
By using the Rot.Sol./CFT method, the central charge of the dual CFT is also calculated as \cite{Chen:2012yd}
\be
c_{\psi}=6\,q^3\,.
\ee
One can obtain the same result by using the thermodynamical method (\ref{unigen}) for the generic (non-extremal) solution.

\section{Further examples} \label{s3}
In this section, we show that the thermodynamical method to obtain the central charges of the dual CFTs is also valid for some other black hole and black ring solutions. Specifically, we consider the Myers-Perry black hole and double rotating dipole black ring.

\subsection{The Myers-Perry (MP) black hole} 
The Mp black hole \cite{MP} describes a rotating black hole in space-time with dimensions $D\ge5$. In $D=5$, the line element of this solution is \cite{MP,Hawking-Hartle}
\bea
ds^2\!\!&=&\! -\frac {\Delta}{{\rho}^{2}} \left( {dt}-a \sin^2\theta d\phi - \,b \cos^2 \theta d\psi \right) ^2\!+{\frac {{\rho}^{2}{{dr}}^{2}}{\Delta}}+{\rho}^{2}{d\theta }^{2}+ \frac {\sin^2 \theta}{{\rho}^{2}} \bigl[ a{dt}- \left( {r}^2\!+{a}^{2}
 \right) d\phi  \bigr] ^{2} \nn \\ &+&\!\!{\frac { \cos^2 \! \theta}{{\rho}^{2}}\! \left[ b{dt}-\! \left( {r}^{2}\!\!+\!
{b}^{2} \right)\! d\psi  \right] ^{2}}\!\!+\!{\frac {1}{{r}^{2}{\rho}^{2}}\! \left[ ab{dt}-b\! \left( {r}^{2}\!\!+\!{a}^{2} \right)\! \sin^2\! \theta d\phi -a \left( {r}^{2}\!\!+\!{b}^{2} \right)\! \cos^2\! \theta d\psi  \right] ^{2}} \!\!\!,
\eea
where $a$ and $b$ are two rotation parameters and
\be
\Delta=\frac{1}{r^2}(r^2 \!+ a^2)(r^2 + b^2) - 2M, \qquad \rho^2 = r^2 + a^2\cos^2\!\theta + b^2\sin^2\!\theta\,,
\ee
also $\frac{3\pi M}{4G_5}$ is the ADM mass of the black hole. The entropy and temperature of the outer and inner horizons and the angular momenta of the solution are
\be \label{sjmp}
S_{\pm}=\frac{\pi^2 (r_{\pm}^2 \!+ a^2) (r_{\pm}^2\!+ b^2)}{2\, r_{\pm}},\quad T_{\pm}=\pm \frac{r_{\pm}^4 - a^2 b^2}{2\pi r_{\pm} (r_{\pm}^2\! +\! a^2)(r_{\pm}^2 \!+\! b^2)}, \quad J_{\phi}= \frac{\pi M a}{2},\quad J_{\psi}= \frac{\pi M b}{2},
\ee
where $r_+$ and $r_-$ are the outer and inner horizons that satisfy $\Delta\!\big|_{r=r_+}\!\!\!\!=0$ and $\Delta\!\big|_{r=r_-}\!\!\!\!=0$. Therefore, it is possible to find $M$ and $b$ in terms of $r_+,r_-$ and $a$ as
\be \label{mb}
M=\frac{(r_+^2+a^2)(r_-^2+a^2)}{2a^2}\,, \qquad b=\frac{r_+r_-}{a}\,.
\ee
Inserting (\ref{mb}) into (\ref{sjmp}), it is easy to check that $T_+S_+=T_-S_-$ which means that the area (entropy) product is universal. One can also find the universality relation of the MP black hole as 
\be \label{epmp}
S_+S_-=4\pi^2J_{\phi}J_{\psi}\,.
\ee
In the appendix (\ref{ap1}) we review the ASG formalism to find the central charges of the dual CFTs. Applying this formalism, the central charges of the dual CFTs to the near horizon of the extremal MP black hole can be computed as \cite{Lu:2008jk,Ghodsi:2013soa}
\be \label{mp-cc}
c_{\phi}=\frac{3\,\pi}{2} \left( a+b \right) ^{2}b=6J_\psi\,,\quad \qquad c_{\psi}=\frac{3\,\pi}{2} \left( a+b \right) ^{2}a=6J_\phi\,.
\ee
Taking into account the entropy product \eqref{epmp}, it is also possible to find these central charges from the thermodynamical method (\ref{unigen}) as
\be
c_\phi=6\,\frac{\partial}{\partial J_\phi}(J_\phi J_\psi)=6J_\psi\,, \qquad \quad c_\psi=6\,\frac{\partial}{\partial J_\psi}(J_\phi J_\psi)=6J_\phi\,,
\ee
which is in complete agreement with the results obtained in (\ref{mp-cc}).

\subsection{The double rotating dipole black ring (DRDBR)}
The DRDBR is a generalization of Pomeransky-Senkov (neutral double rotating) black ring \cite{Pomeransky:2006bd}  including magnetic dipole charge and dilatonic field, which is introduced in \cite{Chen:2012kd}. In the ring coordinates, the solution is
\bea \label{2sdr}
ds^2_5=\!&-&\!\!\!\left[\frac{H(y,x)^3}{K(x,y)^2H(x,y)}\right]^{\frac13}\!\left(d t+\omega_1\,d\psi+\omega_2\,d\phi\right)^2+\frac{2R^2}{(x-y)^2}\left[K(x,y)H(x,y)^2\right]^{\frac13}\nn \\
&\times&\!\!\left(\frac{F(x,y)\left(d\psi+\omega_3\,d\phi\right)^2}{H(x,y)H(y,x)}-\frac{G(x)G(y)\,d\phi^2}{F(x,y)}+\frac{1}{\Phi\Psi}\bigg[\frac{d x^2}{G(x)}-\frac{d y^2}{G(y)}\bigg]\right),
\eea
where the coordinates lie in ranges $0\leq \phi, \psi\leq 2\pi$ and $-1\leq x \leq 1, -\infty <y < -1$ with infinity located at $x=y=-1$. There are some functions and terms in  metric \eqref{2sdr} that are defined as
\bea
G(x)&=&(1-x^2)(1+cx)\,,\nn\\
K(x,y)&=&-a^2(1+b)\big[bx^2(1+cy)^2+(c+x)^2\big]+\big[b(1+cy)-1-cx\big]^2+bc^2(1-xy)^2,\nn\\
\omega_1&=&\frac{R(1+b)(1+y)J(x,y)}{H(y,x)}\left[\frac{2a(a+c)}{\Phi\Psi}\right]^{\frac12},\nn\\
\omega_2&=&\frac{R c(1+b)(1-x^2)\big[(1+cy)(a+ab+by)-c-y\big]}{H(y,x)}\left[\frac{2ab(a+c)(1-a^2)}{\Phi\Psi}\right]^{\frac12},\nn\\
\omega_3&=&\frac{\sqrt{b(1-a^2)}\,ac(1+b)(x-y)(1-x^2)(1-y^2)}{\Phi\Psi \,F(x,y)}\,,\nn\\
&\times&\left[b(1+cx)(1+cy)(1-b-a^2-a^2b)-(1-c^2)(1-b+a^2+a^2b)\right],\nn\\
\Phi&=&1+a-b+ab\,, \hspace{3cm} \Psi=1-a-b-ab\,.
\eea
The gauge field $A$ and dilaton field $\chi$ of the solution are given by
\be\label{gdil}
A=A_t dt+A_{\phi}\, d\phi+A_{\psi}\, d\psi\,, \qquad e^{-\chi}=\left(\frac{K(x,y)}{H(x,y)}\right)^{\sqrt{2/3}},
\ee
where
\bea\label{gfc}
A_{t}\!\!&=&\!\!- \frac{c(1+b)(1-xy)(x-y)}{K(x,y)}\,[b(a^2-c^2)(1-a^2)]^{1/2},\nn\\
A_{\phi}\!\!&=&\!\!-\frac{R(1+b)(1+x)L(x,y)}{K(x,y)}\left[\frac{2a(a-c)}{\Phi\Psi}\right]^{1/2},\\
A_{\psi}\!\!&=&\!\!-\frac{R c(1\!+\!b)(1\!+\!y)}{K(x,y)}\left[\frac{2ab(a\!-\!c)(1\!-\!a^2)}{\Phi\Psi}\right]^{\!1/2}\!\!\!\!\big[x(1\!-\!y)(1\!+\!c)\Phi+(1\!-\!x)^2(a+ab+bcy+c)\big]\,\nn.
\eea
There are also some messy functions in the above solution, $F(x,y), H(x,y), J(x,y)$ and $ L(x,y)$, for which we refer the reader to \cite{Chen:2012kd}. The DRDBR solution also contains four independent parameters $a, b, c$ and $R$ where $a$ and $b$ control the dipole charge and the $S^2$ rotation, respectively. The scale of the solution is related to $R$ and the size of black ring is characterizes by $c$\,. The outer horizon of (\ref{2sdr}) lies at $y_+=-1/c$\, and there is also an inner horizon located at $y_-=-\infty$\,. The entropy and temperature on the outer horizon and angular momenta of this solution are \cite{Chen:2012kd}
\bea \label{tsj}
T_+\!\!&=&\!\!\frac{1}{8\pi R (1+b)}\,\sqrt{\frac{2\Phi\Psi^3}{a(a+c)(1\!-\!a^2)}}\,,\quad \quad S_+=4\pi^2R^3 c(1+b)\sqrt{\frac{2a(a+c)(1\!-\!a^2)}{\Phi\Psi^3}}\,,\\
J_{\phi}\!&=&\!\!2\pi R^3c(1\!+\!b)\sqrt{\frac{2ab(a\!+\!c)(1\!-\!a^2)}{\Phi\Psi^3}}, \quad J_{\psi}=2\pi R^3(1\!+\!b)[(1\!+c)\Phi+2bc(1\!-\!a)]\sqrt{\frac{a(a\!+c)}{2\Phi\Psi^3}}.\nn
\eea
It is also straightforward to calculate the entropy and temperature of the inner horizon. The result is
\be \label{st}
S_-=4\pi^2R^3bc(1\!+b)\sqrt{\frac{2a(a+c)(1\!-\!a^2)}{\Phi\Psi^3}}\,, \qquad T_-=\frac{1}{8\pi Rb (1\!+b)}\,\sqrt{\frac{2\Phi\Psi^3}{a(a+c)(1\!-\!a^2)}}\,.
\ee
Considering (\ref{tsj}) and (\ref{st}), one can check that $T_+S_+=T_-S_-$\,. In other words the entropy product $S_+S_-$ for the DRDBR solution is universal as
\be \label{univ2sdr}
S_+S_-=4\pi^2 J_{\phi}^2\,.
\ee
The extremal limit of the DRDBR, which is obtained by inserting $a,c \to 0$ and $ b \to 1$, is parametrized by \cite{Chen:2012kd}
\be
a=\frac{c}{2\alpha}\,, \qquad b=1-\frac{c}{\beta}\,,
\ee
where $\alpha, \beta$ are finite parameters. At the extremal limit the Hawking temperature vanishes, however the entropy, mass, charge and angular momenta of the solution remains finite. For instance the angular momenta of  DRDBR at the extremal limit are
\be
J_{\phi}=\frac {2 \pi R^3\alpha \beta^2\sqrt {2+4\,\alpha}}{(\alpha-\beta) \sqrt{\alpha^2-\beta^2} },\qquad J_{\psi}=\frac{\pi {R}^{3} \beta \left( \beta+\alpha+2\alpha\beta\right) \sqrt{2+4\alpha}}{(\alpha-\beta) \sqrt{\alpha^2-\beta^2}}\,.
\ee
Applying the Rot.Sol./CFT analysis, the central charges of the CFTs dual to the near horizon of extremal DRDBR are obtain as \cite{Ghodsi:2014fta}
 \be\label{cc2sdr}
c_{\phi}=\frac {24 \pi R^3\alpha{\beta}^{2}\sqrt {2+4\alpha}}{ \left( \alpha-\beta \right) \sqrt{\alpha^2-\beta^2} }=12 J_\phi\,, \quad \qquad c_{\psi}=0\,.
\ee
Alternatively, one can find the same result by considering (\ref{univ2sdr}) and using the thermodynamical method (\ref{unigen}).

\section{Entropy product, central charges and conical singularity} \label{s4}
It has been shown \cite{Golchin:2019hlg} that in the case of solutions with conical singularity, regardless of the horizon topology of the solutions, the conical characteristic $\kappa$ appears in the entropy product law as $\kappa S_+S_-\sim N$. In this section using the Rot. Sol./CFT analysis, we calculate the central charges of the dual CFTs for some black hole and black ring solutions containing the conical singularity. We find that the central charges receive a contribution from the conical characteristic of the solutions.  However, these results are in complete agreement with the central charges obtained from the thermodynamics method (\ref{unigen}).

\subsection{The charged rotating C-metric}
The C-metric solution \cite{kw}, describes a pair of black holes that are uniformly accelerating away from each other. The generalization of C-metric to a solution that includes rotation and charge (Kerr-Newman black holes), is given by the line element \cite{Astorino:2016ybm}
\bea \label{crcm}
&ds^2 &\!= \frac{1}{(1+ A x r)^2} \bigg[ \frac{f(r) + a^2 h (x)}{r^2 + a^2 x^2 } dt^2  - \frac{r^2 \!+ a^2 x^2 }{f(r)} dr^2 + \frac{r^2 \!+ a^2 x^2 }{h(x)} dx^2 \\
&+&\!\!\! \frac{a^2(1\!-\!x^2)^2 f(r) + (a^2\!+\!r^2)^2 h(x)}{r^2+a^2 x^2} \Delta_\phi^2\, d\phi^2  + 2 \frac{a (1\!-\!x^2) f(r) + a (a^2\!+\!r^2) h(x) }{r^2+a^2 x^2} \Delta_\phi dt d\phi  \bigg],\nn
\eea
where 
\bea \label{fhd}
f(r) &=&(A^2 r^2\! -1) (r-r_+) (r-r_-),\qquad
h(x)=(1 \!-\! x^2) (1+ A x r_+) (1+ A x r_-),\nn\\
\Delta_\phi&=&\left[(1+ A r_+)(1+ Ar_-)\right]^{-1} =\left[1+2mA+A^2(q^2+a^2)\right]^{-1}.
\eea
There are four parameters $m$, $q$, $A$ and $a$ in this solution which are related to the mass,  electric charge, acceleration and angular momentum, respectively. The polar coordinate $x$ and azimuthal coordinate $\phi$ are in the ranges $-1\leq x\leq 1$ and $0\leq \phi\leq 2\pi$. The solution also contains the gauge field $A_\mu$ which is defined  as 
\be
 A_\mu = \left[ \frac{-q r}{r^2 + a^2 x^2} , 0 , 0 ,  \frac{ q r a ( 1-x^2 )}{r^2 + a^2 x^2}\right].
 \ee
It is shown \cite{Hong:2004dm} that, for this solution there are outer and inner horizons at 
\be
 r_\pm = m \pm \sqrt{m^2-q^2-a^2}\,\,.
 \ee
The electric charge, mass, angular momentum and Hawking temperature  are \cite{Astorino:2016ybm,Golchin:2019hlg}
\be \label{qmjcm}
Q=q\Delta_\phi\,, \qquad M=m\Delta_\phi\,, \qquad J=a\,m\Delta_\phi^2\,, \qquad T_H=\frac{(r_+\!-\!r_-)\left(1\!-\!A^2r_+^2\right)}{4\pi(r_+^2+a^2)}\,,
\ee
the conical characteristic is also introduced and calculated for this solution in \cite{Golchin:2019hlg}
\be \label{tkcm}
\kappa=\frac{(1\!-\!Ar_+)(1\!-\!Ar_-)}{(1\!+\!Ar_+)(1\!+\!Ar_-)}=\frac {1\!-2mA+A^2 ({q}^2\!+a^2)}{1\!+2mA+{A}^{2} ({q}^2\!+a^2)}\,.
\ee
According to the Rot. Sol./CFT correspondence, in order to calculate the central charge of the dual CFT, one needs to find the near horizon metric of the extremal solution. We remind that at the extremal limit the Hawking temperature vanishes and the inner and outer horizons coincide. So in the case of charged rotating C-metric solution the extremal limit is obtained by setting $m^2=q^2+a^2$, that results in
\be
T_H=0\,, \qquad r_-=r_+=m\,.
\ee
Following the method in appendix (\ref{ap1}), we find the near horizon of the extremal solution as
\be \label{NHEcrcm}
ds^2=\alpha(x)\big[-r^2dt^2+\frac{dr^2}{r^2}\big]+\beta(x)dx^2 +\gamma(x)(d\phi+f^\phi r dt)^2\,,
\ee
where
\bea
\alpha(x)&=&\frac{-r_+^2-a^2x^2}{(1+Axr_+)^2(A^2r_+^2-1)}\,, \qquad \beta(x)=\frac{-r_+^2-a^2x^2}{(x^2-1)(1+Axr_+)^4}\,,\nn\\
\gamma(x)&=&\frac{(1-x^2)(a^2+r_+^2)^2}{(1+Ar_+)^4(r_+^2+a^2x^2)}\,, \qquad f^\phi=\frac{2ar_+(1+Ar_+)}{(1-Ar_+)(a^2+r_+^2)}\,.
\eea
Now using (\ref{ccg}), we obtain the central charge of the CFT dual to the extremal charged rotating C-metric as
\be \label{cccm}
c_\phi=\frac{3}{2\pi}f^\phi\!\int dx d\phi \sqrt{\beta(x) \gamma(x)}=\frac{12am}{(1-Ar_+)^2(1+Ar_+)^2}\,.
\ee
Noticing that $r_-=r_+$ at the extremal limit and comparing the above result with (\ref{fhd}), (\ref{qmjcm}) and (\ref{tkcm}), it is possible to rewrite (\ref{cccm}) as
\be \label{cccmf}
c_\phi=\frac{12\,J}{\kappa}\,,
\ee
note that the conical character $\kappa$ appears in the central charge.

As we mentioned earlier, one may also find the central charge from the thermodynamics of the solution. The entropy and temperature of the inner and outer horizons for the generic (non-extremal) charged rotating C-metric (\ref{crcm}) are \cite{Golchin:2019hlg}
\be
S_\pm=\frac{\pi \Delta_\phi(r_{\pm}^2 + a^2)}{1-A^2 r_{\pm}^2}\,, \qquad T_\pm=\frac{(r_+-r_-)\left(1-A^2r_\pm^2\right)}{4\pi\left(r_\pm^2+a^2\right)}\,,
\ee
it is straightforward to check that $T_+S_+=T_-S_-$ and the entropy product satisfies the universality relation \cite{Golchin:2019hlg}
\be
\kappa S_+S_-=4\pi^2\Big(J^2+\frac{Q^4}{4}\Big)\,.
\ee
 Now, one can read the central charge of the dual CFT from the entropy product of the horizons by using (\ref{unigen}) as
\be
c_{J_\phi}=\frac{6}{4\pi^2}\,\frac{\partial (S_+S_-)}{\partial J }=6\,\frac{\partial}{\partial J} \left[\frac{1}{\kappa}\Big(J^2+\frac{Q^4}{4}\Big)\right]=\frac{12\,J}{\kappa}\,.
\ee
The above result is in complete agreement with (\ref{cccmf}). It means that the thermodynamics method \cite{Chen:2012mh,Chen:2012yd,Chen:2013rb} to find the central charge, remains valid in the case of solutions containing the conical singularity.

\subsection{The unbalanced Pomeransky-Sen'kov black ring}
The Pomeransky-Sen'kov solution \cite{Pomeransky:2006bd} is a neutral five-dimensional double rotating black ring, in which the self-gravity is exactly equal to the centrifugal force due to the rotation in the ring direction. In the absence of this equality, the solution is unbalanced and it contains the conical singularity. The generic unbalanced solution is in the form \cite{Chen:2011jb}
\bea \label{ubr}
ds^2\!&=&\!-\frac{H(y,x)}{H(x,y)}\Big( dt- \omega_\phi d\phi-\omega_\psi d\psi\Big)^2\!+\frac{F(y,x)}{H(y,x)}\,d\phi^2-2\frac{J(x,y)}{H(y,x)}\,d\psi d\phi-\frac{F(x,y)}{H(y,x)}\,d\psi^2 \nonumber\\
&+&\!\frac{2k^2(1-\mu)^2(1-\nu)H(x,y)}{\Phi\Psi(1-\lambda)(1-\mu\nu)(x-y)^2}\left[\frac{d x^2}{G(x)}-\frac{ dy^2}{G(y)}\right],
\eea
where
\be
\Phi\!=\!1-\lambda\mu-\lambda\nu+\mu\nu\,,\quad
\Psi=\mu-\lambda\nu+\mu\nu-\lambda\mu^2, \quad
\Xi=\mu+\lambda\nu-\mu\nu-\lambda\mu^2, \nn
\ee
\bea
 G(x)\!&=&\!\left(1-x^2\right)(1+\mu x)(1+\nu x)\,,\nn\\ 
H(x,y)\!&=&\!\Phi\Psi+\nu(\lambda-\mu)(1+\lambda)\Phi
+\nu{x}^2 y^2 \Psi\Xi+\nu ( \mu+\nu )(\lambda-\mu )(1- \lambda\mu)( 1-\lambda\mu{x}^{2}{y}^{2} ) \nn\\
&+&\!\lambda (\mu+\nu)\big[1-\lambda\mu-\nu (\lambda-\mu)xy\big]\, \big[(1-\lambda\mu ) x+ \nu( \lambda-\mu) y \big]\,,\nn\\
J(x,y)\!&=&\!\frac{2k^2(\mu+\nu)(1-x^2)(1-y^2)\,\sqrt{\nu(\lambda-\mu)(1-\lambda\mu)}}{\Phi(1-\mu\nu)(x-y)}\,\Big[
\Phi\Psi+\nu(\lambda-\mu)(1+\lambda)\Phi\nn\\
&-&\!\nu\Psi\Xi xy+\nu (\mu+\nu )(\lambda-\mu )(1-\lambda\mu ) ( 1+\lambda x+\lambda y+\lambda\mu xy)\Big]\,,\nn
\eea
\bea
F(x,y)\!&=&\!\frac{2k^2}{\Phi\mu\nu(1-\mu\nu)(x-y)^2}\Bigg\{G(x)(y^2-1)\,\bigg\{\mu(1-\lambda^2)[\Psi+\nu(\lambda-\mu)(1+\nu)]^2\nn\\
&-&\!(\mu+\nu)(1-\lambda\mu)(1+\nu y) \Big[\Psi\Xi-\lambda\mu(\lambda-\mu)[\Psi+\nu(\lambda-\mu)(1+\nu)]\Big]\bigg\}\nn\\
&+&\!\nu G(y)\bigg\{\!(\lambda\!-\!\mu)(1\!-\!\lambda\mu)\Big[
\lambda(\mu+\nu)^2(1\!-\!\lambda\mu)+[\Psi+\nu(\lambda\!-\!\mu)(1\!+\nu)](\mu\!+\nu-\mu\nu x)x\Big]\nn\\ &+&\![\Psi\Xi+\lambda\mu\Phi(\Phi-1)(\Phi-\Psi+\Xi) 
][1+(\mu+\nu)x]x^2\nn\\
\!&+&\!\!\mu\nu\Phi[\Psi\Xi-\lambda\mu(\mu+\nu)(\lambda-\mu)(1-\lambda\mu)]x^4
\bigg\}\Bigg\}.
\eea
The solution contains four parameters. Three dimensionless parameters $\nu, \mu$ and $\lambda$ are in the ranges $0\leq \nu \leq \mu \leq \lambda<1$. The forth one is $k>0$ with the dimension of length and it sets the scale of the solution. The spatial coordinates in this solution lie in the ranges $0\leq \phi, \psi \leq 2\pi$, $-1\leq x\leq 1$ and $-\infty <y\leq -1$ where the infinity is at $x=y=-1$. The unbalanced metric (\ref{ubr}) has two horizons which are the roots of $G(y)$. The outer horizon is $y_+=-1/\mu$ and the inner one is located at $y_-=-1/\nu$\,. The entropy, Hawking temperature, mass and angular momenta of this solution are \cite{Chen:2011jb} 
\bea \label{ubrth}
S_+\!&=&\!\frac{4\,\pi^2 k^3\Xi(\mu+\nu)(1-\mu)}{(1-\lambda)(1+\mu)}\left(\frac{2\lambda(1+\lambda)(1-\nu)}{(1-\mu\nu)^3\Phi\Psi}\right)^{\!\frac12}\,, \nn\\ T_+\!&=&\!\frac{(\mu-\nu)(1-\lambda)(1+\mu)}{8\,\pi k\Xi(\mu+\nu)(1-\mu)}\left(\frac{2(1-\mu\nu)\Phi\Psi}{\lambda(1+\lambda)(1-\nu)}\right)^{\!\frac12}\,,\nn\\
M\!&=&\!\frac{3\pi k^2\!\lambda\Phi(\mu+\nu)(1\!-\mu)}{2\Psi(1\!-\lambda)(1\!-\mu\nu)}\,,\qquad  J_{\phi}=\frac{2\pi k^3(\mu+\nu)(1-\mu)}{(1-\mu\nu)^{\frac32}}\left(\frac{2\nu\lambda(1+\lambda)\Xi}{(1-\lambda)\Phi\Psi}\right)^{\!\frac12},\\
J_{\psi}\!&=&\!\frac{\pi k^3(\mu+\nu)(1\!-\mu)[2\nu(1\!-\!\lambda)(1\!-\mu)+(1\!-\nu)\Phi]}{(1-\lambda)^{\frac32}\,(1-\mu\nu)^{\frac32}\, \Psi^{\frac32}}\left(\frac{2\lambda(\lambda-\mu)(1+\lambda)(1-\lambda\mu)\Xi}{\Phi}\right)^{\!\frac12}.\nn
\eea
The conical characteristic, the entropy and temperature of the inner horizon for this solution are \cite{Golchin:2019hlg} 
\bea \label{ubrk}
\kappa&=&\frac{1+\mu}{1-\mu}\,\sqrt{\frac{(1-\lambda)(1+\nu)\Psi}{(1+\lambda)(1-\nu)\Xi}}\,,\nn\\
S_-&=&\frac{4 \sqrt{2}\, \pi ^2 k^3 \nu (\lambda +1) (\mu -1)^2  (\mu +\nu )}{(\nu +1) \Psi }\bigg[\frac{\lambda  (\nu +1) \Xi }{(\lambda -1) \Phi  (\mu  \nu -1)^3}\bigg]^{\frac12}\,,\nn\\
T_-&=&\frac{(\nu +1)(-\mu+\nu) \Psi}{4 \sqrt{2}\, \pi  k (\lambda +1) (\mu -1)^2 (\mu +\nu)(\mu \nu -1)\nu}\bigg[\frac{(\lambda -1) \Phi  (\mu  \nu -1)^3}{\lambda  (\nu +1) \Xi }\bigg]^{\frac12}\,.
\eea
Note that setting $\kappa=1$, leads to the constraint $\lambda=\frac{2\mu}{1+\mu^2}$ which removes the conical singularity and one can recover the Pomeransky-Sen'kov (balanced) solution \cite{Pomeransky:2006bd}. 

In the ASG analysis, the first step to obtain central charges of the dual CFTs is finding the extremal metric. For this solution, the extremality occurs for $\nu=\mu$, where the inner and outer horizons coincide and  the Hawking temperature vanishes. Setting $\nu=\mu$ in (\ref{ubr}), see appendix \ref{ap1} and \cite{Ghodsi:2014fta}, one can find the central charges as 
\be \label{ccubr}
c_\phi=\frac{48\sqrt{2\lambda \mu^3}\,\pi k^3(\mu-1)^2(1+\lambda)^2}{\sqrt{1-2\lambda \mu+\mu^2}\,(1+\mu)^4(\lambda-1)^2}\,, \qquad c_\psi=0\,.
\ee
Considering (\ref{ubrth}), (\ref{ubrk}) and taking into account the extremal condition $\nu=\mu$, one can rewrite this result as
\be \label{ccubr1}
c_\phi=\frac{12\,J_\phi}{\kappa}\,, \qquad c_\psi=0\,.
\ee 
Note again the appearance of the conical characteristic $\kappa$ in the central charge. 

It is also possible to find central charges by considering the thermodynamics of the horizons. Noticing (\ref{ubrth}) and (\ref{ubrk}), it is easy to check that $T_+S_+=T_-S_-$ and the universality of the entropy product takes to the form
\be \label{uniubr}
\kappa\,S_-\,S_+=4\,\pi^2J_\phi^2\,.
\ee
At this stage, using the thermodynamics method (\ref{unigen}), it is easy to read the central charges as
\be
c_{J_\phi}=6\,\frac{\partial}{\partial J_\phi} \left[\frac{J_\phi^{\,2}}{\kappa}\right]=\frac{12\,J_\phi}{\kappa}\,,\qquad c_{J_\psi}=6\,\frac{\partial}{\partial J_\psi} \left[\frac{J_\phi^{\,2}}{\kappa}\right]=0\,,
\ee
which is in complete agreement with (\ref{ccubr1}).

\subsection{The dipole black ring}
As the last solution in this subsection, we consider the dipole black ring. This solution is one of the generalizations of neutral single rotating black ring \cite{Emparan:2001wn} that contains magnetic dipole charge $q$. The metric as presented in \cite{dipole}, is
\bea \label{dbr}
ds^2\!&=&\!-\frac{F(y)H(x)}{F(x)H(y)}\Big[dt+R\sqrt{\lambda(\lambda-
\nu)\frac{1+\lambda}{1-\lambda}}\,\frac{1+y}{F(y)}d\psi\Big]^2\\
&+&\!\frac{R^2F(x)H(x)H(y)^2}{(x\!-y)^2} \Big[\frac{G(x)}{F(x)H(x)^3}\,d\phi^2+\frac{dx^2}{G(x)}
-\frac{dy^2}{G(y)}-\frac{G(y)}{F(y)H(y)^3}\,d\psi^2
\Big],\nn
\eea
where the functions $F, G$ and $H$ are
\be
F(\chi)=1+\lambda\chi\,,\qquad G(\chi)=(1-\chi^2)(1+\nu\chi)\,,\qquad
H(\chi)=1-\mu\chi\,.
\ee
This solution is characterized by four parameters $\mu, \nu, \lambda$ and $R$ which are in the ranges $0\leq \mu<1$, $0<\nu \leq \lambda<1$ and $R>0$. In fact $\lambda$ and $\nu$ determine the shape and rotation velocity, $\mu$ controls the dipole charge and $R$ is related to the  size of the ring. 
The black ring (\ref{dbr}) has two horizons: the outer one is at $y=-1/\nu$  and the the inner horizon lies at $y=-\infty$. The Hawking temperature, mass, angular momentum, dipole charge and entropy for this solution are \cite{dipole}
\bea \label{drth}
T_+\!&=&\!\frac{\nu(1+\nu)}{4\pi R(\mu+\nu)^{3/2}}\sqrt{\frac{1-
\lambda}{\lambda(1+\lambda)}}\,,\hspace{11mm} M=\frac{3\pi R^2(1+\mu)^3}{4(1-\nu)}\left[\lambda+\frac{\mu(1-\lambda)}{1+\mu}\right]\,,\nn\\
J_\psi\!&=&\!\frac{\pi R^3(1\!+\!\mu)^{9/2}}{2 (1-\nu)^2} \sqrt{\lambda(\lambda-\nu)(1\!+\!\lambda)}\,,\quad \,\, q=\frac{R(1\!+\!\mu)(2\pi)^{1/3}}{(1-\nu)\sqrt{1-\mu}}\sqrt{\mu(\mu+\nu)(1\!-\!\lambda)}\,,\nn\\
S_+\!&=&\!2\pi^2 R^3\frac{(1+\mu)^3(\mu+\nu)^{3/2}\sqrt{\lambda(1-
\lambda^2)}}{(1-\nu)^2(1+\nu)}\,.
\eea
Moreover, the entropy and temperature of the inner horizon are \cite{Castro:2012av}
\be
S_-=2\pi^2R^3\frac{(1+\mu)^3\mu^{3/2}}{(1-\nu)^2}\sqrt{\lambda(\lambda-\nu)(1\!-\!\lambda^2)}\,,\qquad
T_-=\frac{\nu}{4\pi R }\sqrt{1-\lambda\over \mu^3 (1\!+\!\lambda)(\lambda-\nu)}\,,
\ee
one can also calculate the conical characteristic for this solution as \cite{Golchin:2019hlg}
\be \label{drk}
\kappa=\frac {\nu+1}{\nu-1}\, \sqrt{\frac{(\mu+1)^3 (\lambda-1)}{(\mu-1)^3(\lambda+1)}}\,.
\ee
At the limit $\nu \to 0$\,, the Hawking temperature vanishes and the inner and outer horizons coincide, which means that the solution is extremal. According to the ASG formalism (appendix \ref{ap1}), it is possible to find the central charges of the dual CFTs for the dipole black ring as \cite{Chen:2012yd}
\be \label{ccdr}
c_\phi=0\,,\qquad c_\psi=12\,\pi R^3\mu^3(1+\mu)^{\frac32}(1-\lambda)\sqrt{1+\lambda}\,.
\ee
Considering the conserved charges in (\ref{drth}), the conical characteristic (\ref{drk}) and noting the extremality condition $\nu=0$, the central charges (\ref{ccdr}) can be written as
\be \label{ccdr1}
c_\phi=0\,,\qquad c_\psi=6\,\frac{q^3}{\kappa}\,.
\ee
Similar to the previous cases, the central charges receive a contribution from the conical characteristic $\kappa$\,.

For the dipole black ring solution, using (\ref{drth})-(\ref{drk}) one can easily check that $T_+ S_+=T_-S_-$ and the universality of entropy product takes the following form 
\be\label{uni}
\kappa\,S_-\,S_+=4\,\pi^2J_\psi q^3\,,
\ee
 alternatively it is possible to read the central charges from the entropy product. In this case, using (\ref{unigen}) one finds
\be
c_\phi=6\,\frac{\partial}{\partial J_\phi} \left[\frac{J_\psi q^3}{\kappa}\right]=0\,,\qquad c_\psi=6\,\frac{\partial}{\partial J_\psi} \left[\frac{J_\psi q^3}{\kappa}\right] =6\,\frac{q^3}{\kappa}\,,
\ee
which is completely in agreement with (\ref{ccdr}).  It shows that the thermodynamics method (\ref{unigen}) works in the case of solutions containing the conic singularity as well.


\section{Conclusions} \label{s5}
In this work using the ASG analysis, we found the central charges of the dual CFTs for the MP black hole and the double rotating dipole black ring. By explicit calculations we also checked that the entropy product for these solutions is universal. In addition, we checked that the central charges of the dual CFTs can be read correctly from the entropy product (\ref{unigen}). In other words, the thermodynamical method (\ref{unigen}) to find central charges, works appropriately in these solutions.

Furthermore, we found the central charges of the dual CFTs, by using the ASG analysis for three solutions that contain the conical singularity: the charged rotating C-metric, the unbalanced Pomeransky-Sen'kov black ring and the dipole black ring. Our results show that the central charges receive a contribution from the conical characteristic $\kappa$ of the solutions.

We also found that in the case of solutions with conical singularity,  it is still possible to read the central charges of the dual CFTs from the entropy product of inner and outer horizons as (\ref{unigen}). Because of the presence of $\kappa$ in the entropy product law for these solutions, the central charges obtained from (\ref{unigen}) are explicitly contain the conical characteristic $\kappa$\,. We found that these central charges are in complete agreement with those obtained from the ASG analysis. In other words, our results show that the thermodynamical method (\ref{unigen}) works properly in the case of solutions with conical singularity.  


\section*{Acknowledgment}
I would like to thank M. M. Sheikh-Jabbari and Kamal Hajian for useful discussions. I also would like to thank the referee for his/her constructive comments.

\appendix 
\section{Near horizon limit and asymptotic symmetry group formalism} \label{ap1}
In this appendix, we will construct a general framework to calculate the central charges of the dual CFTs for the extremal solutions. A five dimensional extremal rotating black hole(ring) solution in the ADM form can be written  as
\be \label{adm}
ds^2=-N^2 dt^2+g_{RR}dR^2+g_{xx}dx^2+g_{ij}(d\phi^i+N^i dt)(d\phi^j+N^j dt),
\ee
where $x$ is the polar angle and $i,j=(\phi,\psi)$. Here we set the location of the degenerate horizon to be at $R=0$. The shift and laps functions in the above metric can be found as
\bea \label{char}
N^\phi=\frac{g_{t\phi}\,g_{\psi\psi}-g_{t\psi}\,g_{\phi\psi}}{g_{\phi\phi}\,g_{\psi\psi}-g_{\phi\psi}^2}\,, \qquad 
N^\psi=\frac{g_{\phi\phi}\,g_{t\psi}-g_{\phi\psi}\,g_{t\phi}}{g_{\phi\phi}\,g_{\psi\psi}-g_{\phi\psi}^2}\,,\qquad 
N^2\!=-g_{tt}+g_{ij}N^i N^j\!.
\eea
In this framework, the first step is finding the near horizon metric. For this purpose, we  expand the quantities around the horizon $R=0$. The result is
\be
N^2=f_1(x)^2 R^2+\mathcal{O} (R^3),  \quad g_{RR}=\frac{f_2(x)^2}{R^2} +\mathcal{O}(R)^{-1}\!, \quad N^i=-\Omega^i+f_3^i R+\mathcal{O}(R)^2\!.
\ee
It is worth mentioning that $N^{\phi}$ and $ N^{\psi}$ are angular velocities of the solution. These terms in the limit of $R\to 0$ give the angular velocities of the horizon which are denoted by $\Omega^i$\,. Now by applying the rescaling \cite{Chen:2011wm} in the form
\be
t \to \frac{f_4}{\varepsilon}\,t \,,\qquad R \to \varepsilon R\,,\qquad   \phi^i \to  \phi^i+\frac{\Omega^i f_4}{\varepsilon}t\,,\qquad f_4=\frac{f_2(x)}{f_1(x)}\,,
\ee
and taking the limit $\varepsilon \to 0\,$, the resulting near horizon metric can be written in the following form
\be \label{NHEM}
ds^2=\alpha(x) \big[-r^2 dt^2+\frac{dr^2}{r^2}\big]+\beta(x)dx^2 +\gamma_{ij}(x)(d\phi^i+f^i r dt)(d\phi^j+f^j r dt)\,,
\ee
where $ f^i=f^i_3 f_4$ for $i=\phi, \psi$\,.

Since the geometry (\ref{NHEM}) has the $SL(2,R)\times U(1)_\phi \times U(1)_\psi$ isometry, we will try to find the central charges of the Virasoro algebras due to the isometry enhancement.  We will follow the asymptotic symmetry group (ASG) approach \cite{Guica:2008mu,BH}; so we need an adequate boundary condition. Let's consider the following boundary condition for the metric fluctuations
 \be
 h_{\mu\nu}\sim\left(\begin{array}{ccccc}
\mathcal{O}(r^2)& \mathcal{O}(1)& \mathcal{O}(1/{r})& \mathcal{O}(r) & \mathcal{O}(r) \\
 &\mathcal{O}({1}/{r^3})& \mathcal{O}({1}/{r^2})&\mathcal{O}({1}/{r})&\mathcal{O}({1}/{r^2})\\
 & &\mathcal{O}({1}/{r})&\mathcal{O}({1}/{r})&\mathcal{O}({1}/{r})\\
 & & &\mathcal{O}(1)&\mathcal{O}(1)\\
 & & & &\mathcal{O}(1/r)
 \end{array}\right)\label{h0}\,.
 \ee
Now it will be easy to show that the general diffeomorphism which preserves this boundary condition is
\be
\zeta=\!\big[C\!+\mathcal{O}(\frac{1}{r^3})\big]\rd_t\!+\!\big[-r\epsilon'(\phi)+\mathcal{O}(1)\big]\rd_r+
 \mathcal{O}(\frac{1}{r})\rd_\theta
+\!\big[\epsilon(\phi)+\mathcal{O}(\frac{1}{r^2})\big]\rd_{\phi}+\mathcal{O}(\frac{1}{r^2})\rd_{\psi}\,,
 \ee
 where C is a constant and $\epsilon(\phi)$ is a  smooth periodic function of $\phi$ and a prime denotes the derivative with respect to $\phi$. Using the basis $\epsilon_n(\phi)=-e^{-in\phi}$ for the function $\epsilon(\phi)$, the generators of asymptotic symmetry group are
 \be
 \zeta_n=e^{-in\phi}\big(-in\, r \rd_r-\rd_{\phi}\big)\,,\label{g-br}
 \ee
which satisfy the Witt Algebra  $[\zeta_m , \zeta_n]=-i(m-n)\zeta_{m+n}$. There is also another set of ASG generators corresponding to the rotation in the $\psi$ direction. Note that in this case  the boundary condition is the same as (\ref{h0}) where $\phi$ and $\psi$ coordinates are exchanged. These generators make another Witt algebra and are given by
\be
\zeta_n=e^{-in\psi}\big(-in\, r \rd_r-\rd_{\psi}\big)\,.\label{g-br2}  
\ee
 Each generator of  diffeomorphism has a conserved charge.  The conserved charges associated to the diffeomorphisms
 (\ref{g-br}) or (\ref{g-br2}) are defined by \cite{Guica:2008mu}
 \be
 Q_\zeta=\frac{1}{8\pi}\int_{\rd\Sigma}k_{\zeta}[h,g],\label{Q}
 \ee
 where $\rd\Sigma$ is a spatial surface at infinity (boundary) and
 \be
 k_\zeta[h,g]=\frac{1}{2}[\zeta_\nu\!\nabla\!_\mu h-\zeta_\nu\!\nabla\!_\sigma h_\mu^{~\sigma}+\zeta_\sigma\!\nabla_\nu h_\mu^{~\sigma}+
\frac{h}{2}\nabla\!_\nu\zeta_\mu-h_\nu^{~\sigma}\nabla\!_\sigma\zeta_\mu
+\frac{1}{2}h_{\nu\sigma}(\nabla\!_\mu\zeta^\sigma+\nabla^\sigma\zeta_\mu)]*(dx^\mu\wedge
 dx^\nu),
 \ee
 in which  ``$*$'' denotes the Hodge dual in five dimensions. In the Brown-Henneaux  approach \cite{BH}
  the central charge is given by
 \be
 \frac{1}{8\pi}\int_{\rd\Sigma}k_{\zeta_m}[\mathcal{L}_{\zeta_n}g,g]
 =-\frac{i}{12}c(m^3- m)\delta_{{m+n},0}\,,\label{Qc}
 \ee
 where $\mathcal{L}_{\zeta_n}g$ is the Lie derivative of the metric (\ref{NHEM}) with respect to ASG Killing vectors (\ref{g-br}) or (\ref{g-br2}).
By doing the calculations, one can find the central charges of the dual CFTs as
\be \label{ccg}
c_i=\frac{3}{2\pi}f^i\!\int dx d\phi d\psi \sqrt{\beta(x) \gamma},\qquad i=\phi,\psi\,,
\ee
where $\gamma=det\gamma_{ij}(x)$.



 \end{document}